\documentclass[12pt]{article}

\usepackage{amssymb,amsmath,amsfonts,eurosym,ulem,graphicx,caption,color,setspace,sectsty,comment,footmisc,caption,natbib,pdflscape,subfigure,array,hyperref}

\usepackage{algorithm, algpseudocode, mathtools}
\usepackage[driver=pdftex]{geometry}
\usepackage{lscape}
\usepackage{hyperref}

\newcolumntype{L}[1]{>{\raggedright\let\newline\\arraybackslash\hspace{0pt}}m{#1}}
\newcolumntype{C}[1]{>{\centering\let\newline\\arraybackslash\hspace{0pt}}m{#1}}
\newcolumntype{R}[1]{>{\raggedleft\let\newline\\arraybackslash\hspace{0pt}}m{#1}}

\begin{document}

\begin{titlepage}
\title{Bayesian ranking and selection with applications to field studies, economic mobility, and forecasting}
\author{Dillon Bowen\thanks{Wharton School of Business, University of Pennsylvania, dsbowen@wharton.upenn.edu}}
\date{\today}
\maketitle
\begin{abstract}
\noindent Decision-making often involves ranking and selection. For example, to assemble a team of political forecasters, we might begin by narrowing our choice set to the candidates we are confident rank among the top 10\% in forecasting ability. Unfortunately, we do not know each candidate's true ability but observe a noisy estimate of it. This paper develops new Bayesian algorithms to rank and select candidates based on noisy estimates. Using simulations based on empirical data, we show that our algorithms often outperform frequentist ranking and selection algorithms. Our Bayesian ranking algorithms yield shorter rank confidence intervals while maintaining approximately correct coverage. Our Bayesian selection algorithms select more candidates while maintaining correct error rates. We apply our ranking and selection procedures to field experiments, economic mobility, forecasting, and similar problems. Finally, we implement our ranking and selection techniques in a user-friendly Python package \citep{Bowen2022multiple} documented here: \url{https://dsbowen-conditional-inference.readthedocs.io/en/latest/}\footnote{Code available here: \url{https://gitlab.com/dsbowen/selection}}.\\
\vspace{0in}\\

\bigskip
\end{abstract}
\setcounter{page}{0}
\thispagestyle{empty}
\end{titlepage}
\pagebreak \newpage


In 2011, the United States Intelligence Community launched a tournament ``to dramatically enhance the accuracy, precision, and timeliness of intelligence forecasts for a broad range of event types, through the development of advanced techniques that elicit, weight, and combine the judgments of many intelligence analysts" \citep{matheny2015aggregative}. The eventual winner of this tournament was a team called the \textit{Good Judgment Project} \citep{tetlock2016superforecasting, mellers2014psychological}. In its first year, the Good Judgment Project gathered forecasts from thousands of participants about geopolitical events ranging from a possible civil war in Syria to Palestinian membership in the United Nations. The Intelligence Community scored the teams based on forecasting accuracy. If they correctly predicted the winner of the Nicaraguan presidential election, for example, they would receive a good score.

The Good Judgment Project set out to assemble a team of ``superforecasters" who ranked in the top 2\% in forecasting ability for the second year of the competition. Of course, the Good Judgment Project did not know each participant's true forecasting ability. Instead, it had to select forecasters using noisy estimates of forecasting ability based on each participant's performance in the tournament's first year.

Ranking and selection problems, like those faced by the Good Judgment Project, are ubiquitous in decision-making. Jack Welch famously fired the bottom 10\% of his workforce each year when he ran General Electric. The Organization for Economic Cooperation and Development regularly ranks countries, for example, by math, science, and reading scores so countries can model their education systems after those that rank near the top \citep{schleicher2019pisa}. Economists rank teachers by the value they add to their students' education \citep{chetty2014measuringa, chetty2014measuringb, hanushek2011economic} and United States neighborhoods by the economic opportunity they afford their children \citep{chetty2018impactsa, chetty2018impactsb}. Social scientists run large-scale field experiments to compare many treatments or interventions simultaneously and select the most effective for further research and implementation \citep{milkman2021megastudies}.

Researchers and practitioners often perform ranking and selection using point estimates alone. For example, the Good Judgment Project ranked forecasters according to point estimates of their forecasting accuracy and selected those whose point estimates fell in the top 2\%. Unfortunately, this procedure has high error rates \citep{gu2020invidious}. Suppose that all the Good Judgment Project's forecasters were about equally talented. By chance, some would appear much better than others. In a world where the variance in estimated forecasting ability is almost entirely due to chance, the Good Judgment Project's false discovery rate would be $1 - .02 = 98\%$. That is, 98\% of their ``superforecasters" would not rank among the top 2\% in forecasting ability.

We want statistical procedures that account for noise when ranking and selecting candidates to address this issue. When ranking candidates, we want \textit{rank confidence intervals} (RCIs) with correct coverage. For example, an RCI might suggest that a particular forecaster is between the 10th- and 50th-best with 80\% confidence. RCIs have correct \textit{marginal} coverage if each covers its individual candidate's true rank with the desired confidence. RCIs have correct \textit{simultaneous} coverage if all simultaneously cover their candidate's true rank with the desired confidence. Recent research proposed frequentist procedures for obtaining RCIs with correct marginal and simultaneous coverage \citep{mogstad2020inference}.

When selecting candidates, we want to select sets of candidates with correct false discovery or family-wise error rates. For example, we might select 40 forecasters that rank in the top 10\% in forecasting ability with a false discovery rate of 5\%. The selection set has a correct false discovery rate if fewer than $0.05 * 40 = 2$ selected forecasters are not genuinely in the top 10\%. Controlling the false discovery rate allows us to claim that 95\% of the forecasters on our team are in the top 10\% in forecasting ability.

Alternatively, we might want to control the family-wise error rate. For example, we might choose 20 forecasters that rank in the top 10\% with a family-wise error rate of 5\%. The selection set has a correct family-wise error rate if there is at least a $1 - .05 = 95\%$ chance that all 20 are genuinely in the top 10\%. Controlling the family-wise error rate allows us to claim that we are 95\% confident that all of the forecasters on our team are in the top 10\% in forecasting ability.

Given a false discovery or family-wise error rate, we usually want to select the largest possible set of candidates. For example, General Electric probably wants to fire as many under-performing employees as possible.

Recent research proved that a Bayesian selection procedure based on posterior tail probabilities has correct false discovery rates \citep{gu2020invidious}. However, this procedure does not control family-wise error rates. Researchers have also proposed using RCIs to perform selection \citep{mogstad2022comment}. For example, if we want to select forecasters in the top 50 with 80\% confidence, we can select those whose 80\% RCI lies entirely between one and 50. Selection based on marginal RCIs (i.e., rank confidence intervals with correct marginal coverage) results in correct false discovery rates. Similarly, selection based on simultaneous RCIs results in correct family-wise error rates. Therefore, we can use \citet{mogstad2020inference}'s frequentist procedures for marginal and simultaneous RCIs to select candidates with correct false discovery and family-wise error rates, respectively \citep{mogstad2022comment}.

This paper develops new Bayesian algorithms that improve upon existing ranking and selection methods. First, we develop new Bayesian algorithms to produce marginal and simultaneous RCIs. Using simulations based on nine datasets, we show that our Bayesian procedures yield substantially shorter RCIs than their frequentist counterparts with approximately correct marginal and simultaneous coverage.

Because we can use RCIs for selection, our Bayesian RCIs can select candidates with correct false discovery and family-wise error rates. Additionally, because our Bayesian RCIs are shorter than frequentist RCIs, Bayesian RCIs select more candidates than frequentist RCIs. Still, it is possible to do better. Indeed, our most significant theoretical contribution is a ``direct" Bayesian selection algorithm which selects more candidates than simultaneous RCIs while maintaining correct family-wise error rates.

Our most significant practical contribution is a user-friendly statistics package implementing our Bayesian ranking and selection algorithms \citep{Bowen2022multiple} documented here: \url{https://dsbowen-conditional-inference.readthedocs.io/en/latest/}. We also provide template notebooks and preconfigured virtual environments. These allow practitioners to upload a \verb|csv| file of their conventional estimates (e.g., ordinary least squares or instrumental variables estimates) and click ``run" to analyze their data without downloading any software or writing a single line of code.

The rest of our paper is structured as follows. Section \ref{sec:ranking} describes frequentist and Bayesian ranking procedures and analyzes their performance using simulations and empirical illustrations. We find that Bayesian RCIs have approximately correct coverage and are substantially shorter than their frequentist counterparts. Section \ref{sec:selection} introduces our direct Bayesian selection algorithm. It then analyzes various selection procedures' performance using simulations and empirical illustrations. We find that Bayesian selection procedures - especially our direct Bayesian selection algorithm - have correct error rates but select many more candidates than frequentist procedures. Finally, section \ref{sec:conclusion} concludes.

\section{Rank confidence intervals} \label{sec:ranking}

Suppose we have $K$ candidates with values $\mu_1,...,\mu_K$. For example, $\mu_k$ might be forecaster $k$'s true forecasting ability. We do not observe each candidate's true forecasting ability but instead a noisy estimate of it $Y_k$. For example, $Y_k$ might by forecaster $k$'s average Brier score from the tournament's first year. In many applications, the estimates come from a joint normal distribution $Y \sim \mathcal{N}(\mu, \Sigma)$, motivated by the central limit theorem. We assume the covariance matrix $\Sigma$ is known, although, in practice, we usually replace it with a consistent estimate. Define the rank of candidate $k$ as

\[
	r_k \coloneqq 1 + \sum_{j=1}^K \mathbf{1}\{\mu_k < \mu_j\}.
\]

A set of rank confidence intervals $R = \{R_1,...,R_K\}$ has correct marginal coverage at level $\alpha$ if each rank confidence interval covers its candidate's true rank with probability $1 - \alpha$,

\[
	Pr\{r_k \in R_k\} \geq 1 - \alpha \quad \forall k \in \{1,...,K\}.
\]

RCIs with correct marginal coverage are \textit{marginal rank confidence intervals} (MRCIs).

We say that a set of rank confidence intervals has correct simultaneous coverage at level $\alpha$ if all the rank confidence intervals simultaneous cover their candidate's true rank with probability $1 - \alpha$,

\[
	Pr\big\{r_k \in R_k \quad \forall k \in \{1,...,K\}\big\} \geq 1 - \alpha.
\]

RCIs with correct simultaneous coverage are \textit{simultaneous rank confidence intervals} (SRCIs).

\subsection{Frequentist and Bayesian construction}

\citet{mogstad2020inference} develops an impressive frequentist procedure for constructing RCIs. To construct an MRCI for candidate $k$, we form simultaneous confidence intervals for the pairwise differences $\mu_k - \mu_1,...,\mu_k-\mu_K$ and then count how many intervals lie entirely above or below zero. Suppose there is one candidate $j$ for whom the 95\% confidence interval around pairwise difference $\mu_k - \mu_j$ lies entirely below zero and $K-5$ candidates for whom the confidence interval around the pairwise difference lies entirely above zero. Then, candidate $k$ ranks between second and fifth with 95\% probability.

Constructing SRCIs follows a similar procedure. We begin by forming simultaneous confidence intervals around all pairwise differences $\mu_j - \mu_k \quad \forall j, k$. Then, as for the MRCIs, we count how many confidence intervals around the pairwise difference between candidate $k$ and another candidate $\mu_k - \mu_1,...,\mu_k - \mu_K$ lie entirely above or below zero.

We now introduce new Bayesian algorithms for RCIs. Instead of treating the values $\mu$ as fixed, we treat them as a random draw from a prior distribution $\mu \sim \mathcal{G}$. We can use Bayesian estimators to obtain a joint posterior distribution $\mu | Y \sim \mathcal{F}$ \citep{james1992estimation, stein1956inadmissibility, cai2021nonparametric, brown2009nonparametric}. Given the posterior, the marginal coverage probability for a RCI $R_k$ is

\[
	Pr\{r_k \in R_k | Y\} = \int_{\mu \in \mathbb{R}^K} \mathbf{1}\big\{
		1 + \sum_{j=1}^K \mathbf{1}\{\mu_k < \mu_j\} \in R_k
	\big\}
	f(\mu | Y) d \mu,
\]

where $f$ is the joint posterior density. In practice, we estimate coverage probabilities by sampling from the joint posterior distribution.

Suppose we want our RCIs to be convex. A naive algorithm might compute an MRCI $R_k$ by searching over all possible convex RCIs and selecting the shortest with correct coverage. This naive algorithm takes quadratic time for each candidate $k$. Therefore, constructing MRCIs for all candidates using the naive algorithm takes cubic time. Algorithm \ref{alg:marginal_conf_int} presents a more efficient method for constructing MRCIs\footnote{The results we present later further restrict RCIs to include each candidate's estimated rank. For example, if a particular forecaster came in 50th place during the first year, we restrict their RCI to include 50. This modification does not affect the algorithm's time complexity.}. This algorithm will find the same RCI as the naive algorithm but in linear time for each candidate (so, quadratic time for all candidates).

\begin{algorithm}
\caption{Bayesian marginal rank confidence intervals}
\label{alg:marginal_conf_int}
\begin{algorithmic}
    \State Given a marginal coverage probability $\alpha$
    \State $l^*_k \gets 1$
    \State $u^*_k \gets K$
    \State $l_k \gets K$
    \State $u_k \gets K$
    \State Estimate $Pr\{r_k = j | Y\} \quad \forall j \in \{1,...,K\}$ using joint posterior samples
    \State $P \gets Pr\{r_k = K\}$
    \While{$l_k > 0$}
        \While{$P < 1 - \alpha$}
            \If{$l_k = 1$}
                \State break
            \EndIf
            \State $l_k \gets l_k - 1$
            \State $P \gets P + Pr\{r_k = l_k | Y\}$
        \EndWhile
        \If{$u_k - l_k < u^*_k - l^*_k$}
            \State $l^*_k \gets l_k$
            \State $u^*_k \gets u_k$
        \EndIf
        \State $u_k \gets u_k - 1$
        \State $P \gets P - Pr\{r_k = u_k | Y\}$
    \EndWhile
    \State \textbf{Result.} Marginal rank confidence interval $R_k = [l^*_k, u^*_k]$
\end{algorithmic}
\end{algorithm}

Similarly, the simultaneous coverage probability for a set of rank confidence intervals $\{R_1,...,R_K\}$ is

\[
\begin{split}
  Pr\big\{r_k \in R_k \quad \forall k \in \{1,...,K\} | Y \big\} \\
	= \int_{\mu \in \mathbb{R}^K} \mathbf{1}\big\{
		1 + \sum_{j=1}^K \mathbf{1}\{\mu_k < \mu_j\} \in R_k
		\quad \forall k \in \{1,...,K\}
	\big\}
	f(\mu | Y) d \mu.  
\end{split}
\]

A naive algorithm might construct SRCIs by searching over all possible sets of convex RCIs and selecting those that are shortest on average with correct simultaneous coverage. However, this is an exponential time algorithm. Algorithm \ref{alg:simultaneous_conf_int} presents a quadratic time algorithm for SRCIs, although its RCIs will be longer than the naive algorithm's in expectation.

\begin{algorithm}
\caption{Bayesian simultaneous rank confidence intervals}
\label{alg:simultaneous_conf_int}
\begin{algorithmic}
    \State Given a simultaneous coverage probability $\alpha$
    \State $R \gets \{R_1,...,R_K\} = \{[l_1, u_1],...,[l_K, u_K]\}$ using algorithm \ref{alg:marginal_conf_int}
    \State $P \gets Pr\big\{r_k \in R_k \quad \forall k \in \{1,...,K\} | Y \big\}$
    \While{$P < 1 - \alpha$}
        \State $m \gets \arg \max_k Pr\{r_k = l_k - 1 | Y\}$
        \State $n \gets \arg \max_k Pr\{r_k = u_k + 1| Y\}$
        \If{$Pr\{r_m = l_m - 1 | Y \} > Pr\{r_n = u_n + 1 | Y \}$}
            \State $P \gets Pr\{r_k \in R_k \quad \forall k \neq m, r_m \in R_m \text{ or } r_m = l_m - 1 | Y\}$
            \State $R_m \gets [l_m - 1, u_m]$
        \Else
            \State $P \gets Pr\{r_k \in R_k \quad \forall k \neq n, r_n \in R_n \text{ or } r_n = u_n + 1 | Y\}$
            \State $R_n \gets [l_n, u_n + 1]$
        \EndIf
    \EndWhile
    \State \textbf{Result.} Simultaneous rank confidence intervals $\{R_1,...,R_K\}$
\end{algorithmic}
\end{algorithm}

\subsection{Simulations} \label{sec:ranking_simulations}

Next, we compare \citet{mogstad2020inference}'s frequentist procedures with our Bayesian procedures for constructing RCIs using simulations based on nine datasets. Because our simulation procedure is identical for all datasets, we begin by describing our simulation procedure in detail for a single dataset, then briefly describe our remaining datasets.

\textbf{Superforecasters.} This dataset contains estimates of forecasting accuracy from the first year of the Good Judgment Project \citep{goodjudgment2016data}. Over 1,000 forecasters predicted more than 100 geopolitical events with discrete or categorical outcomes (e.g., ``[From 0-100\%, how likely is it that the] Spanish government generic 10-year bond yields equal or exceed 7\% at any point before 1 September 2012?"). After filtering out forecasters who responded to fewer than half the forecasting questions, our dataset included 842 forecasters. The Good Judgment Project scored each forecast using Brier scores \citep{tetlock2016superforecasting}. We can estimate each forecaster's ability by averaging their Brier scores across the forecasting questions to which they responded\footnote{Technically, Brier scores measure \textit{inaccuracy}, so we use negative Brier scores to rank candidates from most to least accurate.}. So, for each forecaster, we have a point estimate $Y_k$ and standard error $\sigma_k$. By the central limit theorem, the point estimates are approximately normally distributed $Y \sim \mathcal{N}(\mu, \Sigma)$ for $\Sigma_{j,k} = \mathbf{1}\{j = k\} \sigma_k^2$. We aim to rank the forecasters based on their forecasting accuracy.

For each of 1,000 simulations, we began by randomly selecting 200 forecasters to rank. (Running 1,000 simulations using the full dataset of 842 forecasters is computationally prohibitive.) We treated the point estimates $Y$ as the ground truth. Therefore, the ground truth rankings are $r_k = 1 + \sum_{j=1}^K \mathbf{1}\{Y_k < Y_j\}$. We then sampled ``new" estimates $Y' \sim \mathcal{N}(Y, \Sigma)$. Using the new estimates $Y'$ and covariance matrix $\Sigma$, we constructed MRCIs and SRCIs using the frequentist and Bayesian procedures for error rates $\alpha = 0.05, 0.1, 0.2$. For MRCIs, $1-\alpha$ is the desired marginal coverage probability. For SRCIs, $1-\alpha$ is the desired simultaneous coverage probability. Finally, we compared the RCIs in marginal and simultaneous coverage and average length.

\textbf{Movers.} The Moving to Opportunity (or ``Movers") dataset estimates the causal effect on children's outcomes in adulthood of moving to a new commuting zone \citep{chetty2016effects, bergman2019creating}. We aim to rank 50 United States commuting zones based on how they affect the future earnings of children who grow up in them.

\textbf{Opportunity Altas.} The Opportunity Atlas is similar to the Movers dataset. It measures children's outcomes in adulthood based on the commuting zone in which they grew up \citep{chetty2018opportunity}. The primary difference between the Opportunity Altas and Movers datasets is that the Opportunity Altas dataset is correlational, whereas the Movers dataset is causal. We aim to rank the 100 largest United States commuting zones based on the future earnings of children who grow up in them.

\textbf{PISA.} In 2016, the Organization for Economic Cooperation and Development conducted the Programme for International Student Assessment (PISA) \citep{schleicher2019pisa}. PISA measured 15-year-olds' reading, mathematics, and science skills in 37 countries. We aim to rank countries based on their students' mathematics scores.

\textbf{24-Hour Fitness.} Behavioral science researchers partnered with 24-Hour Fitness to test the effects of 53 behavioral nudges encouraging 60,000 customers to exercise more \citep{milkman2021megastudies}. The treatments involved planning, reminders, microincentives, and other interventions. The researchers defined the treatment effects as the increase in weekly gym visits during a four-week intervention period compared to a control condition. We aim to rank the treatments based on their effects on weekly gym visits.

\textbf{Flu study.} Behavioral science researchers partnered with Penn Medicine and Geisinger Health to test the effects of 19 text messages encouraging 50,000 patients to get a flu vaccine \citep{milkman2021megastudy}. The treatments involved texting patients various reminders. The researchers measured the treatments' effectiveness as the increase in vaccination rates compared to the control group. We aim to rank treatments based on their effects on vaccination rates.

\textbf{State representatives.} Economists called state representatives' offices 40 times in all 50 United States to estimate the probability of reaching a live representative by state \citep{dube2021note}. We aim to rank states based on the probability of reaching a live representative.

\textbf{Diversity in academia.} Business school researchers sent emails to 6,500 professors at top U.S. universities from fictional students interested in applying to a doctoral program \citep{milkman2015happens}. The fictional students' names signalled their race (White, Black, Hispanic, Chinese, or Indian) and gender. In their emails, the fictional students asked the professors for a meeting. The researchers estimated how often professors agreed to meet students in each demographic group (i.e., groups based on race and gender). We aim to rank demographic groups based on the probability that professors agreed to meet them.

\textbf{Discriminatory sentencing.} Criminology researchers studied sentencing patterns among over 800 federal judges \citep{smith2021racial}. They measured how much longer each judge's sentences were for Black defendants than White defendants, controlling for type of crime committed, criminal record, and other observables. For each simulation, we randomly selected 200 judges to analyze. (Using the entire dataset of 800 judges is computationally prohibitive.) We aim to rank judges based on their level of sentencing discrimination.

Our procedures for constructing RCIs will work for any Bayesian estimator. Because parametric Bayesian estimators outperform nonparametric Bayesian estimators for datasets with few candidates \citep{berger2013statistical}, we use a parametric estimator (normal-prior, normal-likelihood fit using maximum likelihood estimation) for datasets with fewer than 30 candidates. Otherwise, we use a nonparametric estimator (Dirac-delta prior fit using expectation maximization) \citep{cai2021nonparametric}.

\begin{figure}
    \centering
    \includegraphics[scale=.5]{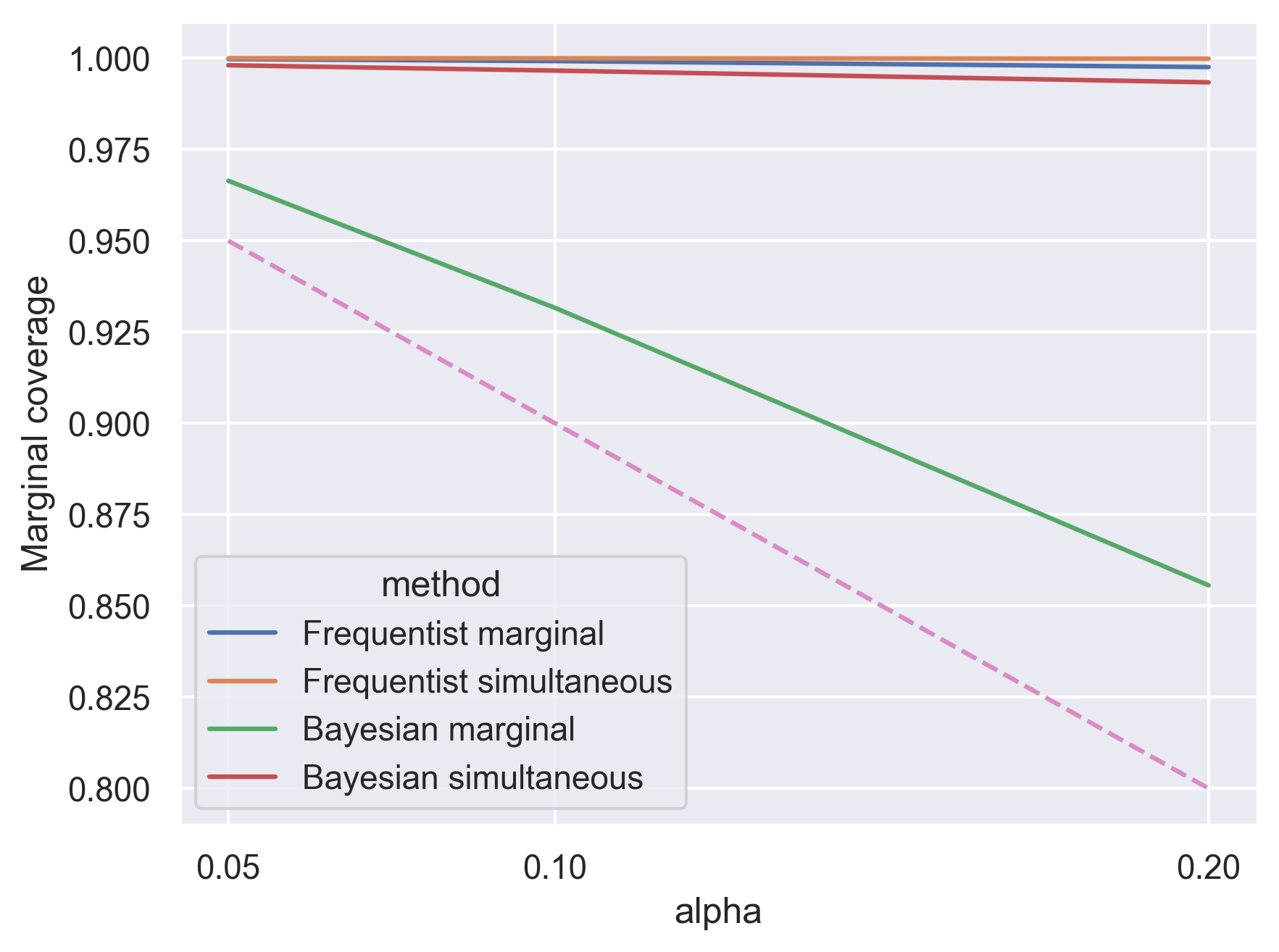}
    \caption{Marginal coverage from 1,000 simulations averaged across all datasets. The coverage is correct if it is above the dashed line.}
    \label{fig:simulations_marginal_coverage}
\end{figure}

Figure \ref{fig:simulations_marginal_coverage} shows that all RCIs (frequentist and Bayesian, marginal and simultaneous) have correct marginal coverage on average across all datasets. Additional analyses (not shown) confirm that all RCIs have correct marginal coverage for each dataset individually.

\begin{figure}
    \centering
    \includegraphics[scale=.5]{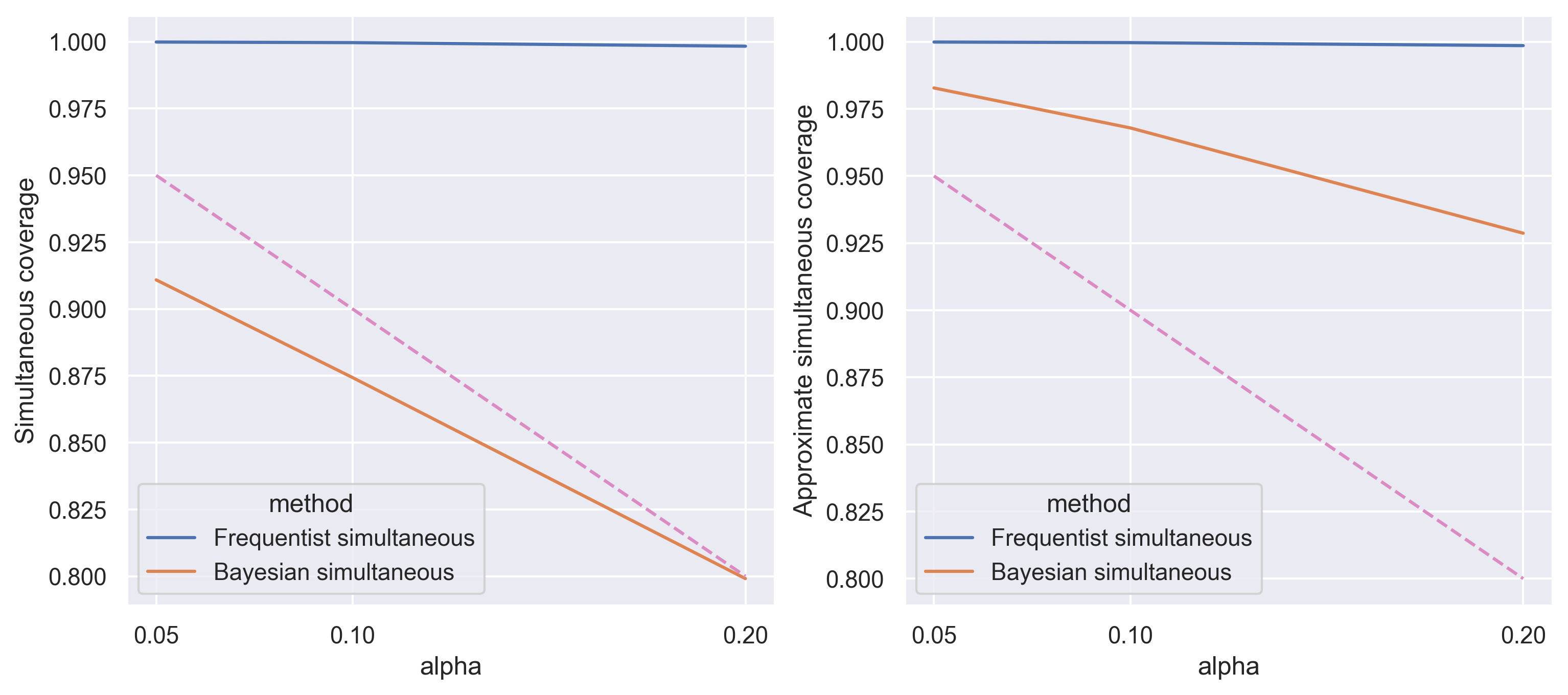}
    \caption{Simultaneous and approximate simultaneous coverage from 1,000 simulations averaged across all datasets. The coverage is correct if it is above the dashed line.}
    \label{fig:simulations_simultaneous_coverage}
\end{figure}

The left panel of Figure \ref{fig:simulations_simultaneous_coverage} shows the simultaneous coverage for the frequentist and Bayesian SRCIs on average across all datasets. Unfortunately, the Bayesian SRCIs have incorrect simultaneous coverage. For example, the 95\% Bayesian SRCIs have only 91\% simultaneous coverage. This is driven primarily by incorrect simultaneous coverage for the Movers, State representatives, and Discriminatory sentencing datasets, for which the 95\% Bayesian SRCIs have only 87\%, 85\%, and 76\% simultaneous coverage, respectively. (The 95\% Bayesian SRCIs have at least 93\% simultaneous coverage for each other dataset individually).

At first, these results appear to reflect poorly on the Bayesian SRCIs. However, the results become more promising if we slightly relax the simultaneous coverage requirement. RCIs have simultaneous coverage if they contain every candidate's true rank. We say that RCIs have \textit{approximate simultaneous coverage} if they contain the true rank of at least 98\% of the candidates. That is, we are willing to tolerate coverage errors for at most one of 50 commuting zones in the Movers dataset, one of 50 states in the State representatives dataset, and four of 200 judges in the Discriminatory sentencing dataset.

The right panel of Figure \ref{fig:simulations_simultaneous_coverage} shows that Bayesian SRCIs have correct approximate simultaneous coverage on average across all datasets. Additional analyses (not shown) confirm that Bayesian SRCIs have correct approximate simultaneous coverage for each dataset individually. For example, the 95\% Bayesian SRCIs for the Movers, State representatives, and Discriminatory sentencing datasets have 99\%, 98\%, and 100\% approximate simultaneous coverage, respectively.

\begin{figure}
    \centering
    \includegraphics[scale=.5]{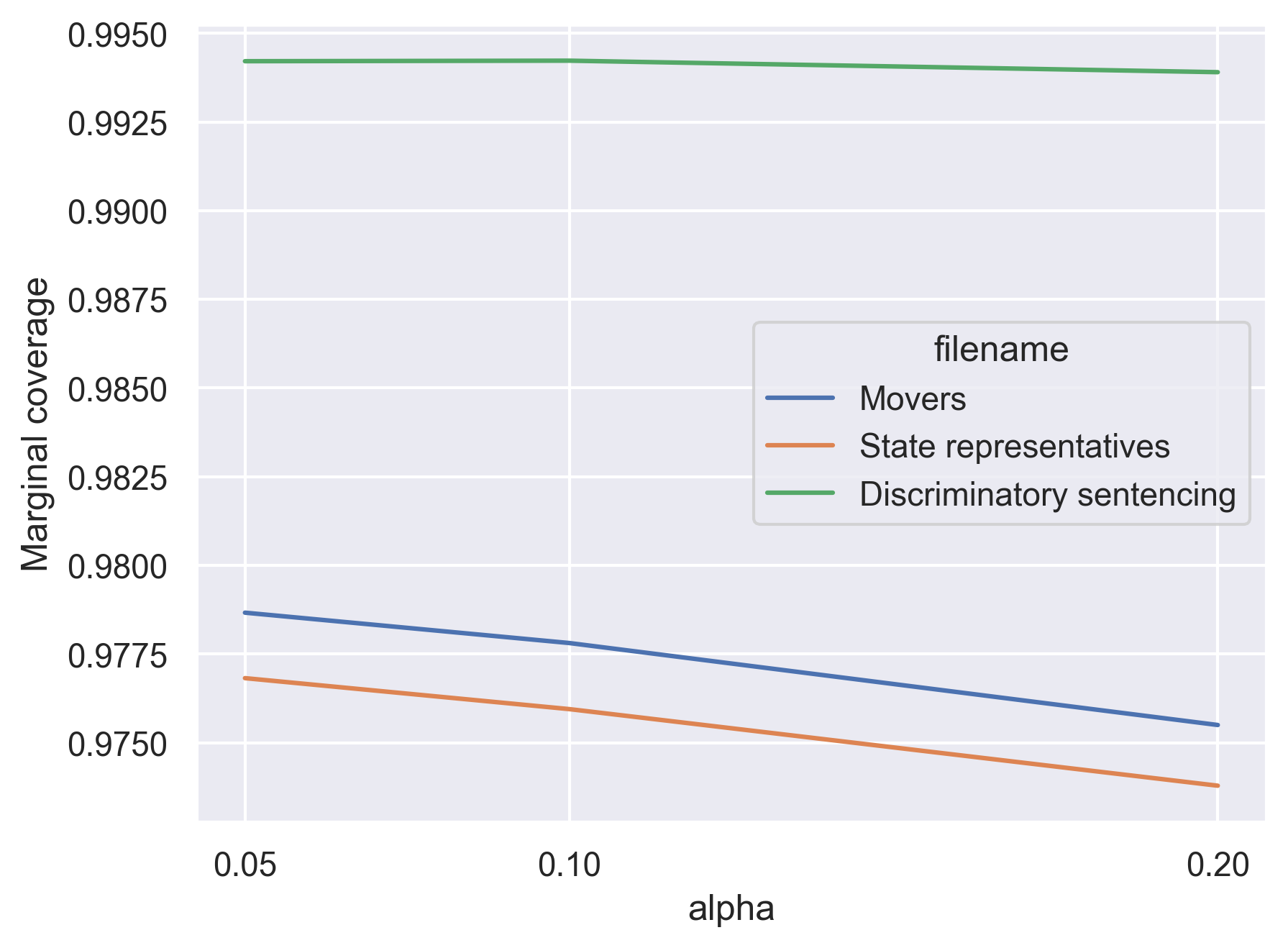}
    \caption{Marginal coverage in simulations where Bayesian SRCIs do not cover the ranks of all candidates.}
    \label{fig:simulations_residual_marginal_coverage}
\end{figure}

Additionally, it is worth asking what the marginal coverage is when the simultaneous coverage is incorrect. For example, the 95\% Bayesian SRCIs have only 76\% simultaneous coverage for the Discriminatory sentencing dataset. What is the average marginal coverage across the 1-0.76=24\% of simulations where the RCIs do not cover all 200 judges? Figure \ref{fig:simulations_residual_marginal_coverage} plots the marginal coverage for the simulations in which the 95\% Bayesian SRCIs do not cover every candidate's rank. In this case, the 95\% Bayesian SRCIs for the Movers, State representatives, and Discriminatory sentencing datasets have 98\%, 98\%, and 99.5\% marginal coverage, respectively. That is, when the 95\% Bayesian SRCIs do not cover every candidate's rank, they cover nearly 49 out of 50 commuting zones in the Mover's dataset, 49 out of 50 states in the State representatives dataset, and 199 out of 200 judges in the Discriminatory sentencing dataset on average.

In sum, Bayesian SRCIs do not have correct simultaneous coverage. However, they have correct \textit{approximate} simultaneous coverage, meaning that the $1-\alpha$ Bayesian SRCIs cover at least 98\% of the candidates' true ranks with probability $1-\alpha$. Additionally, when they do not achieve (exact) simultaneous coverage, Bayesian SRCIs have high marginal coverage and likely fail to cover only a single candidate.

\begin{figure}
    \centering
    \includegraphics[scale=.5]{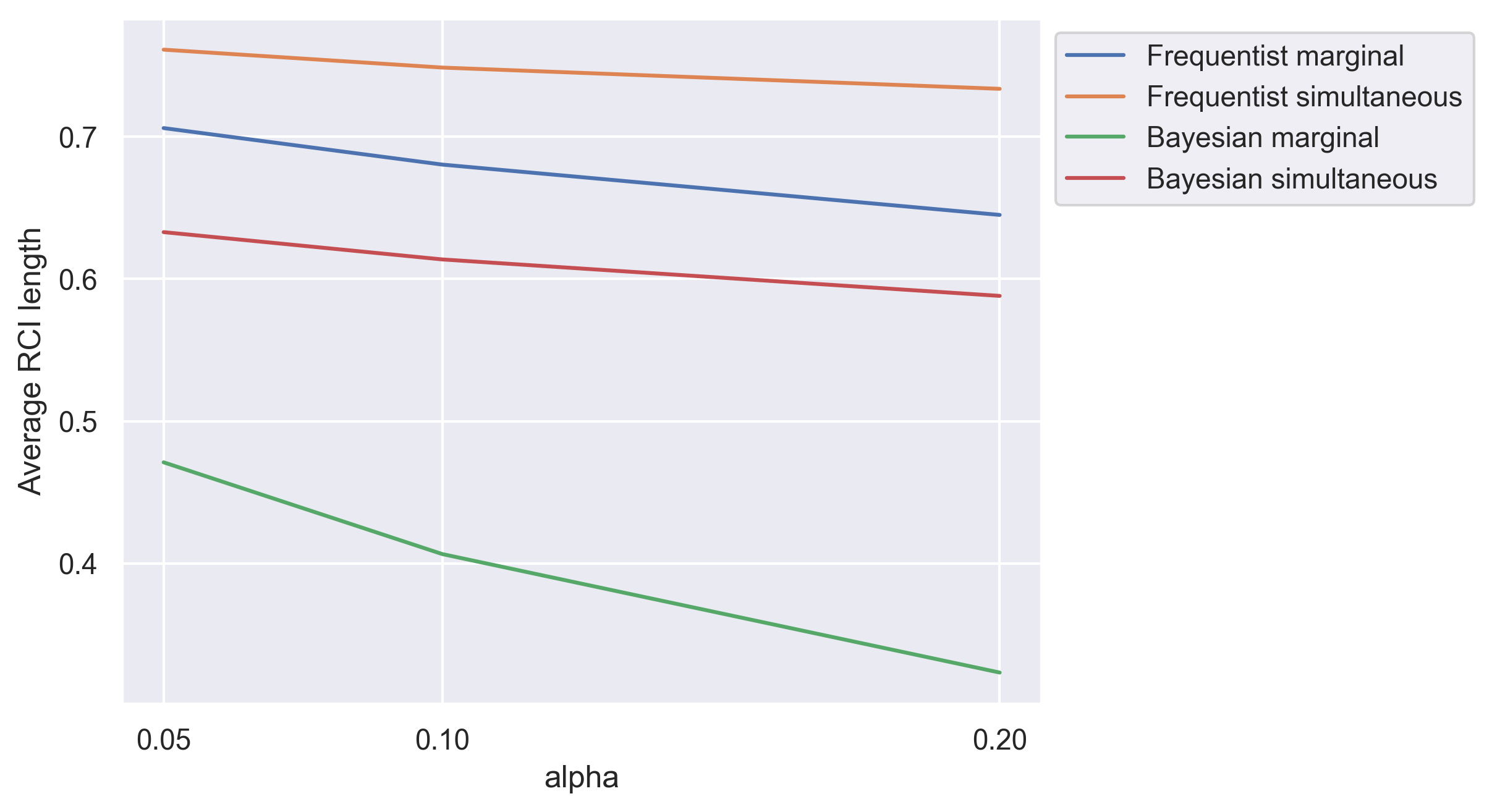}
    \caption{Average rank confidence interval length.}
    \label{fig:simulations_length}
\end{figure}

Having established that Bayesian RCIs have approximately correct coverage, we now ask how long the RCIs are. Figure \ref{fig:simulations_length} plots the average length of the RCIs as a proportion of the candidates in each dataset. For example, if the RCIs were 50 ranks long on average in a dataset of 100 candidates, the average length as a proportion of the candidates is $50 / 100 = 0.5$. 

Bayesian RCIs are substantially shorter than their frequentist counterparts. Bayesian SRCIs are consistently 20\% shorter than frequentist SRCIs, and Bayesian MRCIs are up to 50\% shorter than frequentist MRCIs on average across all datasets. Additional analyses (not shown) confirm that Bayesian RCIs are shorter than their frequentist counterparts for every error rate $\alpha$ in every dataset by similar margins.

\subsection{Empirical illustration}

\begin{figure}
    \centering
    \includegraphics[scale=.5]{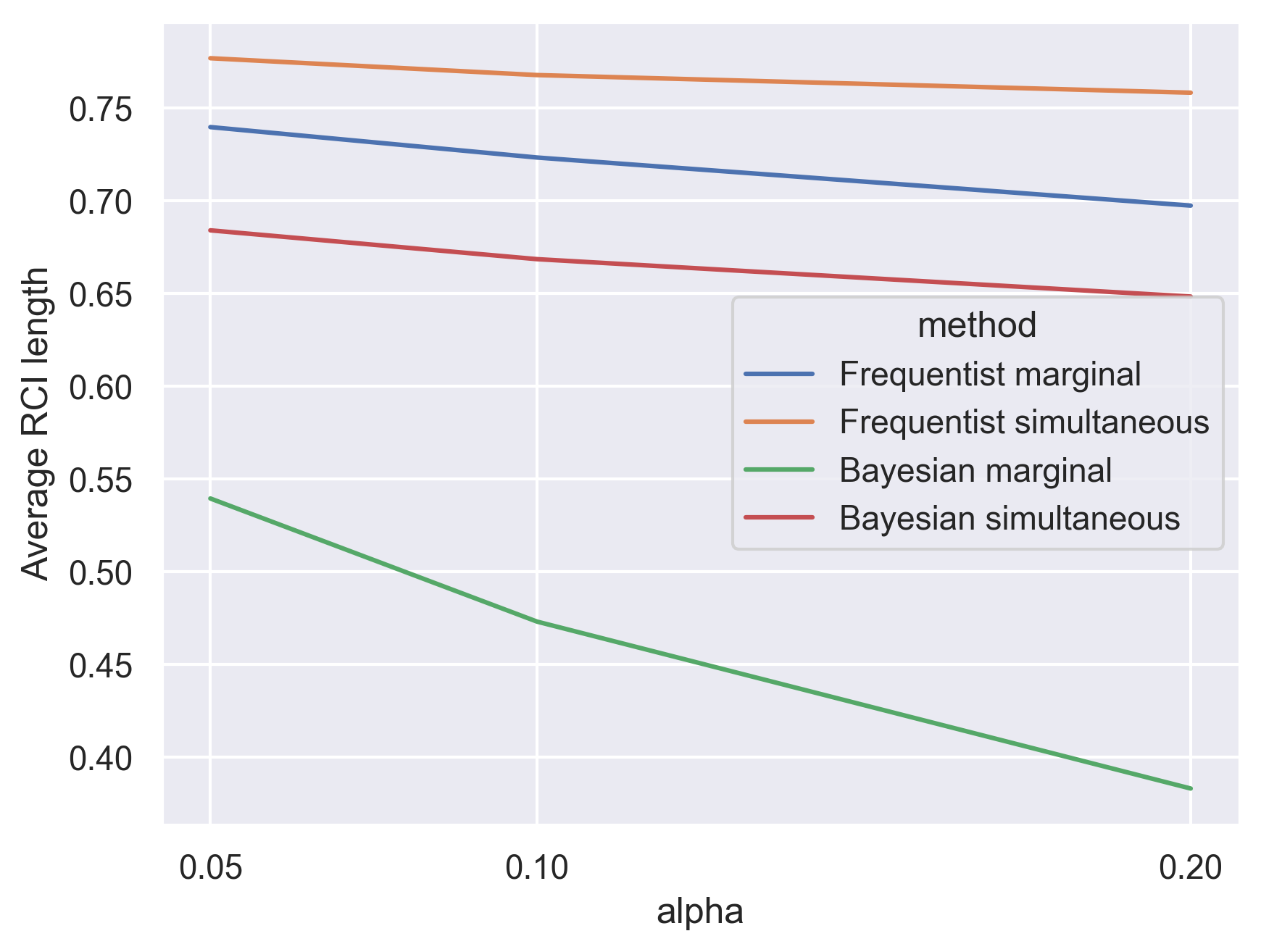}
    \caption{Length of RCIs applied to conventional estimates.}
    \label{fig:empirical_length}
\end{figure}

In our simulations, we used the conventional estimates as the ground truth and sampled new estimates from this distribution. We found that Bayesian RCIs are consistently shorter than their frequentist counterparts. Here, we estimate RCIs using the conventional estimates themselves. Figure \ref{fig:empirical_length} plots the average length of the RCIs across all candidates and datasets. Consistent with our simulation results, the Bayesian RCIs are shorter than their frequentist counterparts.

\begin{landscape}
\begin{figure}
    \centering
    \includegraphics[scale=.4]{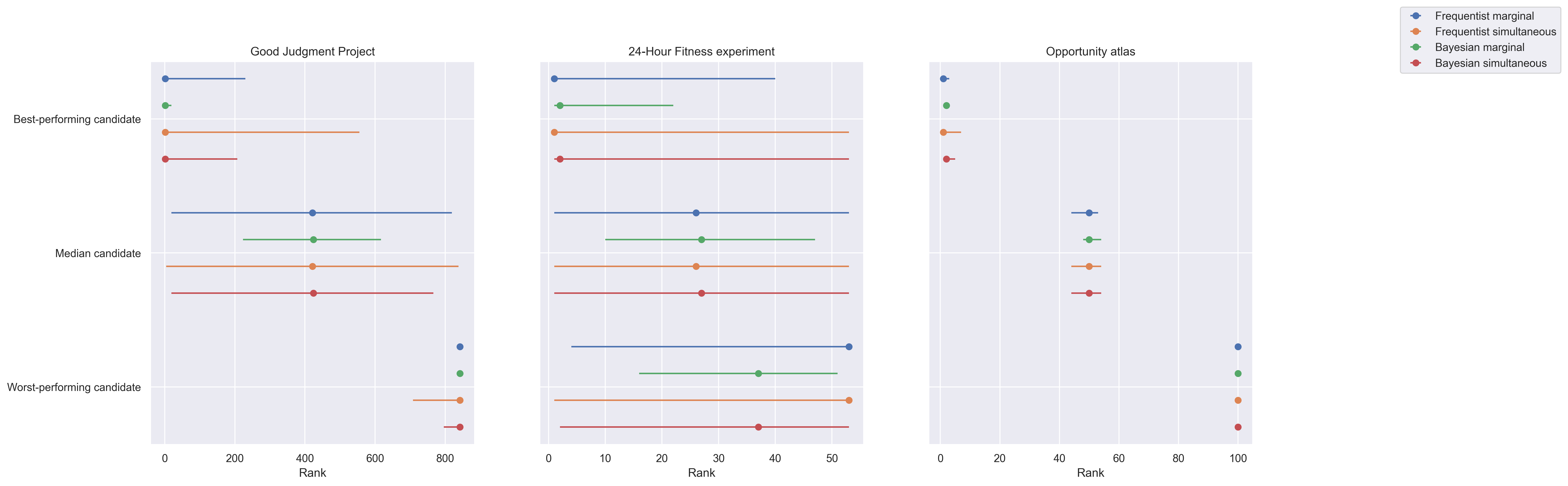}
    \caption{80\% RCIs for the Good Judgment Project, 24-Hour Fitness experiment, and Opportunity Atlas. Note that frequentist and Bayesian models may disagree about the point estimate of a candidate's rank, which is why the blue and orange (frequentist) dots do not exactly align with the green and red (Bayesian) dots.}
    \label{fig:empirical_ci}
\end{figure}
\end{landscape}

Additionally, Figure \ref{fig:empirical_ci} shows the 80\% RCIs for the best-performing, worst-performing, and median candidates in three datasets (the Good Judgment Project, 24-Hour Fitness, and Opportunity Atlas datasets). These results suggest substantial uncertainty for the Good Judgment Project and 24-Hour Fitness datasets. For example, according to Bayesian SRCIs, we cannot be even 80\% confident that the best-performing forecaster in the Good Judgment Project was among the top 200 of 842 forecasters in the tournament. Additionally, the median forecaster (who ranked 421st according to conventional estimates) could be one of the best or worst forecasters. However, it is easier to identify the worst forecasters, as shown by the short Bayesian SRCI for the forecaster who performed worst according to conventional estimates.

Similarly, many of the Bayesian SRCIs for the 24-Hour Fitness experiment span nearly the entire 53 behavioral treatments. By contrast, the RCIs for the Opportunity Atlas are relatively narrow, suggesting that we can confidently rank commuting zones by their children's future income.

\section{Selection} \label{sec:selection}

Selection is closely related to ranking. We often want to select a set of candidates that we are confident rank near the top. Formally, suppose we want to select the top $r^*$ candidates. Let $S \coloneqq \{k : r_k \leq r^*\}$ be the set of truly best candidates. Because we do not know each candidate's true rank, the set of candidates we select is a random set $\hat{S}$.

We say that our selection set has a correct marginal false discovery rate (mFDR) at level $\alpha$ if the percentage of selected candidates ranked below $r^*$ is less than $\alpha$,

\[
    mFDR = \frac{
        \mathbf{E}[|\hat{S} \setminus S|]
    }{
        \mathbf{E}[|\hat{S}|]
    } \leq \alpha.
\]

As the number of candidates increases, the marginal false discovery rate approaches the usual false discovery rate \citep{genovese2002operating},

\[
    FDR = \mathbf{E}\left[
        \frac{
        	|\hat{S} \setminus S|
        }{
        	|\hat{S}|
   	}
    \right].
\]

We say that our selection set has a correct family-wise error rate at level $\alpha$ if all the selected candidates rank above $r^*$ with probability $1 - \alpha$,

\[
	FWER = Pr\{\hat{S} \setminus S = \emptyset\} \geq 1 - \alpha.
\]

Controlling the false discovery or family-wise error rate, we usually prefer to identify as many top candidates as possible. For example, the Good Judgment Project probably prefers to identify as many superforecasters as possible. That is, we want to maximize $|\hat{S}|$ without inflating the false discovery or family-wise error rate.

\subsection{Frequentist and Bayesian construction}

We can easily use ranking procedures to perform selection. For example, if we want to select forecasters in the top $r^*$ with 80\% confidence, we can simply select those whose 80\% RCI lies entirely between one and $r^*$,

\[
	\hat{S} = \{k : R_k \subseteq [1, r^*]\}.
\]

Asserting that candidate $k$ ranks in the top $r^*$ is a directional claim, so one-tailed RCIs are more appropriate than two-tailed RCIs for selection. Selection based on MRCIs leads to correct false discovery rates, whereas selection based on SRCIs leads to correct family-wise error rates \citep{mogstad2022comment}. Because Bayesian RCIs are shorter than frequentist RCIs, we expect that Bayesian RCIs will select more candidates than frequentist RCIs.

However, direct selection algorithms may outperform selection based on RCIs. \citet{mogstad2020inference} propose a direct frequentist selection algorithm that controls the family-wise error rate. Their direct selection algorithm selects more candidates than frequentist SRCIs. Unfortunately, their algorithm takes exponential time and is infeasible in many practical applications.

In the same spirit, we consider direct Bayesian selection algorithms. We start by expressing the false discovery rate of a selection set $\hat{S}$ given the posterior density. The false discovery rate is,

\[
    \mathbf{E}\left[\frac{|\hat{S} \setminus S|}{|\hat{S}|} \Big| Y\right]
    = \frac{1}{|\hat{S}|} \sum_{k \in \hat{S}} \int_{\mu_k \in \mathbb{R}} \mathbf{1}\big\{
        1 + \sum_{j=1}^K \mathbf{1}\{\mu_k < \mu_j\} > r^*
    \big\} f(\mu_k|Y) d\mu_k.
\]

As we did for Bayesian RCIs, we approximate the false discovery rate by sampling from the joint posterior distribution.

A seminal paper on Bayesian selection provides a selection procedure guaranteed to control the marginal false discovery rate based on posterior tail probabilities \citep{gu2020invidious}. Similarly, Algorithm \ref{alg:selection_fdr} is a simple selection procedure based on samples from the joint posterior distribution. We call this the \textit{Bayesian FDR direct} method because it is a direct Bayesian selection method designed to control the false discovery rate. This algorithm begins with an empty set. According to Bayesian estimates, we then select candidates one at a time in order of how likely they are to rank above $r^*$. We stop when we have reached a given false discovery rate, as estimated by the Bayesian posterior.

\begin{algorithm}
\caption{Bayesian selection to control the false discovery rate}
\label{alg:selection_fdr}
\begin{algorithmic}
    \State Given marginal posterior probabilities that candidate $k$ is one of the $r^*$-best $Pr\{r_k \leq r^* | Y\}$
    \State Given a desired false discovery rate $\alpha$
    \State $\hat{S} \gets \emptyset$
    \State $\Omega \gets \{1,...,K\}$
    \State $FDR \gets 0$
    \While{$FDR < \alpha$}
        \State $j \gets \arg \max_{k \in \Omega} Pr\{r_k \leq r^* | Y\}$
        \State $FDR \gets \frac{|\hat{S}| * FDR + 1 - Pr\{r_j \leq r^* | Y\}}{|\hat{S}| + 1}$
        \If{$FDR \geq \alpha$}
            \State break        
        \EndIf
        \State $\hat{S} \gets \hat{S} \cup \{j\}$
        \State $\Omega \gets \Omega \setminus \{j\}$
    \EndWhile
    \State \textbf{Result.} Selected set of candidates $\hat{S}$
\end{algorithmic}
\end{algorithm}

Similarly, the family-wise error rate of a selection set is,

\[
    Pr\{|\hat{S} \setminus S| > 0 | Y\}
    = 1 - \int_{\mu \in \mathbb{R}^K} \mathbf{1}\big\{
        1 + \sum_{j=1}^K \mathbf{1}\{\mu_k < \mu_j\} \leq r^* \quad \forall k \in \hat{S}
    \big\} f(\mu|Y) d\mu.
\]

\citet{gu2020invidious}'s Bayesian selection procedure was not designed to have correct family-wise error rates. To address this issue, we designed Algorithm \ref{alg:selection_fwer} to control the family-wise error rate when performing Bayesian selection. We call this the \textit{Bayesian FWER direct} method because we expect it to control the family-wise error rate. This  algorithm begins with the set of all candidates. We then reject candidates one at a time in order of how likely they are to rank below $r^*$ according to Bayesian estimates (more precisely, we reject the candidate that minimally increases the estimated family-wise error rate). We stop when we have $1-\alpha$ confidence that we have rejected all those who rank below $r^*$, as estimated by the Bayesian posterior. Therefore, we can have $1-\alpha$ confidence that all the remaining candidates rank above $r^*$.

\begin{algorithm}
\caption{Bayesian selection to control the family-wise error rate}
\label{alg:selection_fwer}
\begin{algorithmic}
    \State Given a desired family-wise error rate $\alpha$
    \State $\hat{S} \gets \{1,...,K\}$
    \State $\Omega \gets \emptyset$
    \State $FWER \gets 1$
    \While{$FWER > \alpha$}
        \State $j \gets \arg \max_{k \in \hat{S}} Pr\{r_k > r^* | Y, \sum_{i \in \Omega} \mathbf{1}\{r_i > r^*\} < K - r^*\}$
        \State $\hat{S} \gets \hat{S} \setminus \{j\}$
        \State $\Omega \gets \Omega \cup \{j\}$
        \State $FWER \gets 1 - Pr\{\sum_{k \in \Omega} \mathbf{1}\{r_k > r^*\} \geq K - r^* | Y\}$
    \EndWhile
    \State \textbf{Result.} Selected set of candidates $\hat{S}$
\end{algorithmic}
\end{algorithm}

\subsection{Simulations}

So far, we have considered six selection algorithms. We have four based on RCIs (marginal versus simultaneous and frequentist versus Bayesian) and two direct Bayesian approaches. We now compare these algorithms using simulations with the same datasets and sampling procedure as Section \ref{sec:ranking_simulations}. After sampling ``new" estimates $Y'$, we apply our selection algorithms to select candidates at various quantiles $q$ (i.e., the top 10\%, 20\%, 80\%, or 90\%) with error rates $\alpha = 0.05, 0.1, 0.2$. Here, $\alpha$ is the desired false discovery or family-wise error rate.

Figure \ref{fig:fdr} shows the average marginal false discovery rate across datasets. As expected, every selection procedure has correct false discovery rates. Additional analyses (not shown) suggest that every procedure also has approximately correct false discovery rates for each quantile and dataset \footnote{For some parameter combinations, there is too much noise to accurately estimate the false discovery rate. For example, the estimated false discovery rate using frequentist RCIs for the Diversity in academia dataset when selecting the top 10\% of candidates with a desired false discovery rate of 5\% is 33\%. This is because it selected only six candidates in all 1,000 simulations, and two happened to be false discoveries. Clearly, frequentist RCIs did not select enough candidates across the 1,000 simulations to accurately estimate the false discovery rate.}.

Figure \ref{fig:fwer} shows the average family-wise error rate across datasets for the three procedures that control the family-wise error rate (the frequentist and Bayesian SRCI methods and the Bayesian FWER direct method). As expected, all of these procedures have correct family-wise error rates on average. Additional results (not shown) confirm that these procedures also have approximately correct family-wise error rates for each quantile and dataset.

Figure \ref{fig:simulations_n_selected} shows the average proportion of candidates selected by each procedure across all datasets. Among the procedures that control the false discovery rate (the frequentist and Bayesian MRCI methods and the Bayesian FDR direct method), the Bayesian FDR direct method selects the most candidates. Notably, both Bayesian methods select many more candidates than the frequentist method. Among the procedures that control the family-wise error rate, the Bayesian FWER direct method selects the most candidates by a wide margin. Additional results (not shown) confirm that these patterns hold for each dataset.

\begin{landscape}
\begin{figure}
    \centering
    \includegraphics[scale=.4]{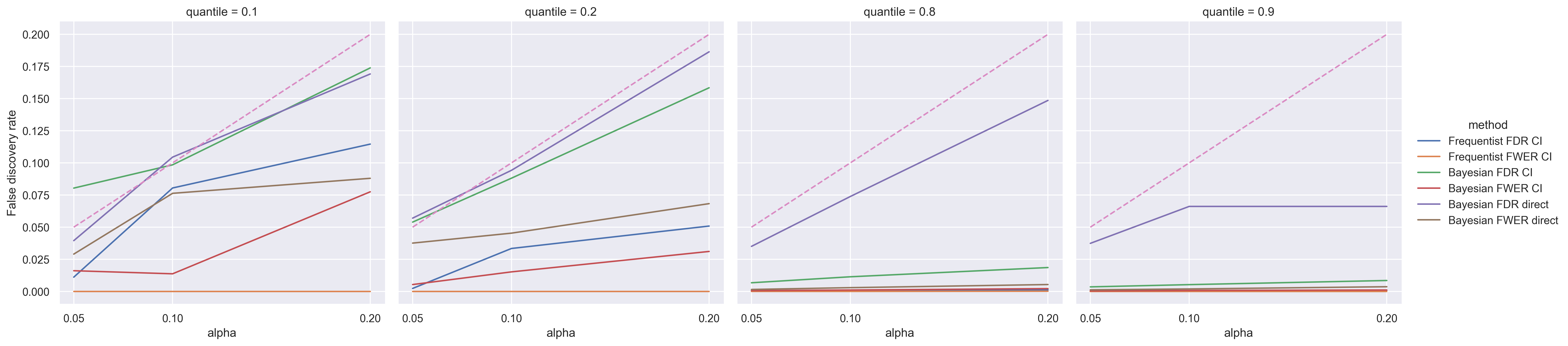}
    \caption{Average marginal false discovery rates across datasets for 1,000 simulations. For comparability, we label the RCI methods according to the error rate they control for. For example, selection based on frequentist SRCIs is labeled \textit{frequentist FWER CI} to highlight that selection based on SRCIs controls the family-wise error rate (FWER). The false discovery rate is correct if it is below the dashed line.}
    \label{fig:fdr}
\end{figure}

\begin{figure}
    \centering
    \includegraphics[scale=.4]{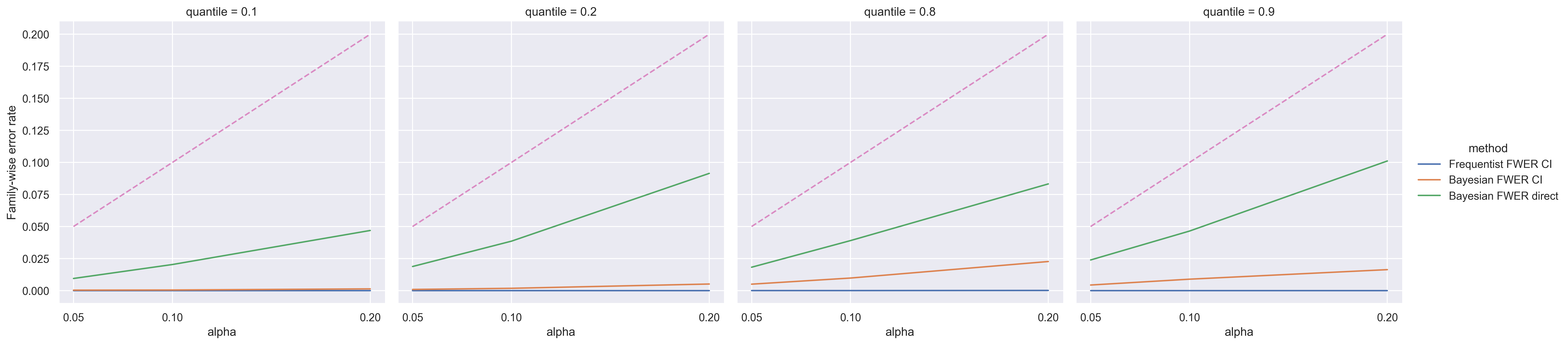}
    \caption{Average family-wise error rates across datasets for 1,000 simulations. The family-wise error rate is correct if it is below the dashed line.}
    \label{fig:fwer}
\end{figure}

\begin{figure}
    \centering
    \includegraphics[scale=.4]{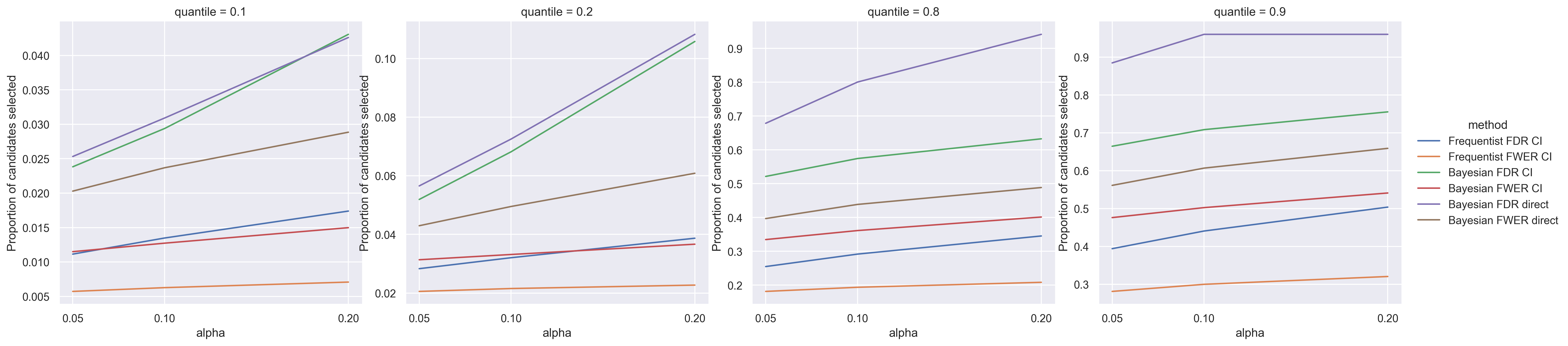}
    \caption{Average proportion of candidates selected across datasets for 1,000 simulations.}
    \label{fig:simulations_n_selected}
\end{figure}
\end{landscape}

\subsection{Empirical illustration} \label{sec:application}

\begin{landscape}
\begin{figure}
    \centering
    \includegraphics[scale=.4]{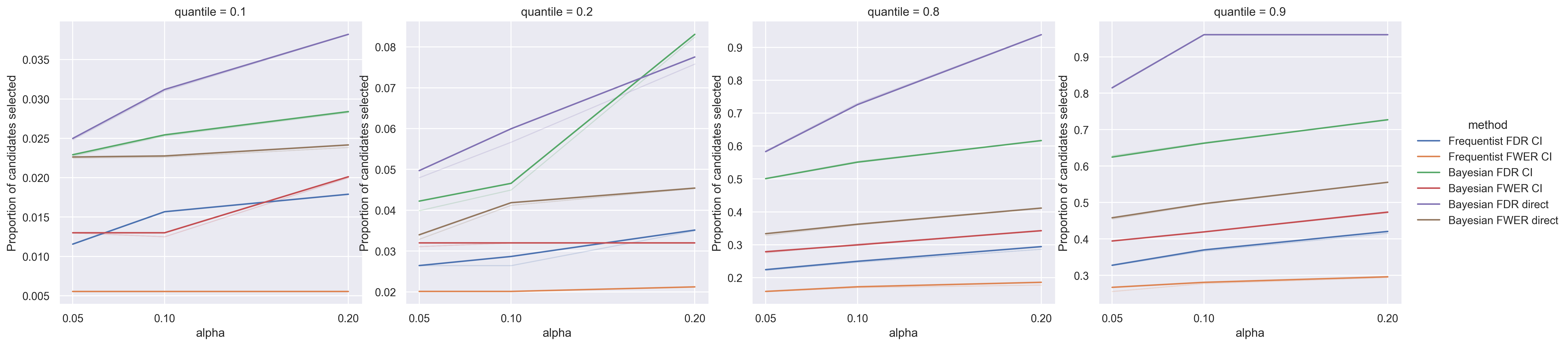}
    \caption{Average proportion of candidates selected across datasets.}
    \label{fig:empirical_n_selected}
\end{figure}
\end{landscape}

In our simulations, we used the conventional estimates as the ground truth and sampled new estimates from this distribution. We found that Bayesian selection procedures, especially the direct procedures, selected more candidates than their frequentist counterparts. Here, we select candidates based on the conventional estimates themselves. Figure \ref{fig:empirical_n_selected} plots the average proportion of selected candidates across all datasets. Consistent with our simulation results, the Bayesian procedures select more candidates than their frequentist counterparts. Additionally, the direct Bayesian selection procedures select more candidates than Bayesian selection based on RCIs.

\begin{figure}
    \centering
    \includegraphics[scale=.5]{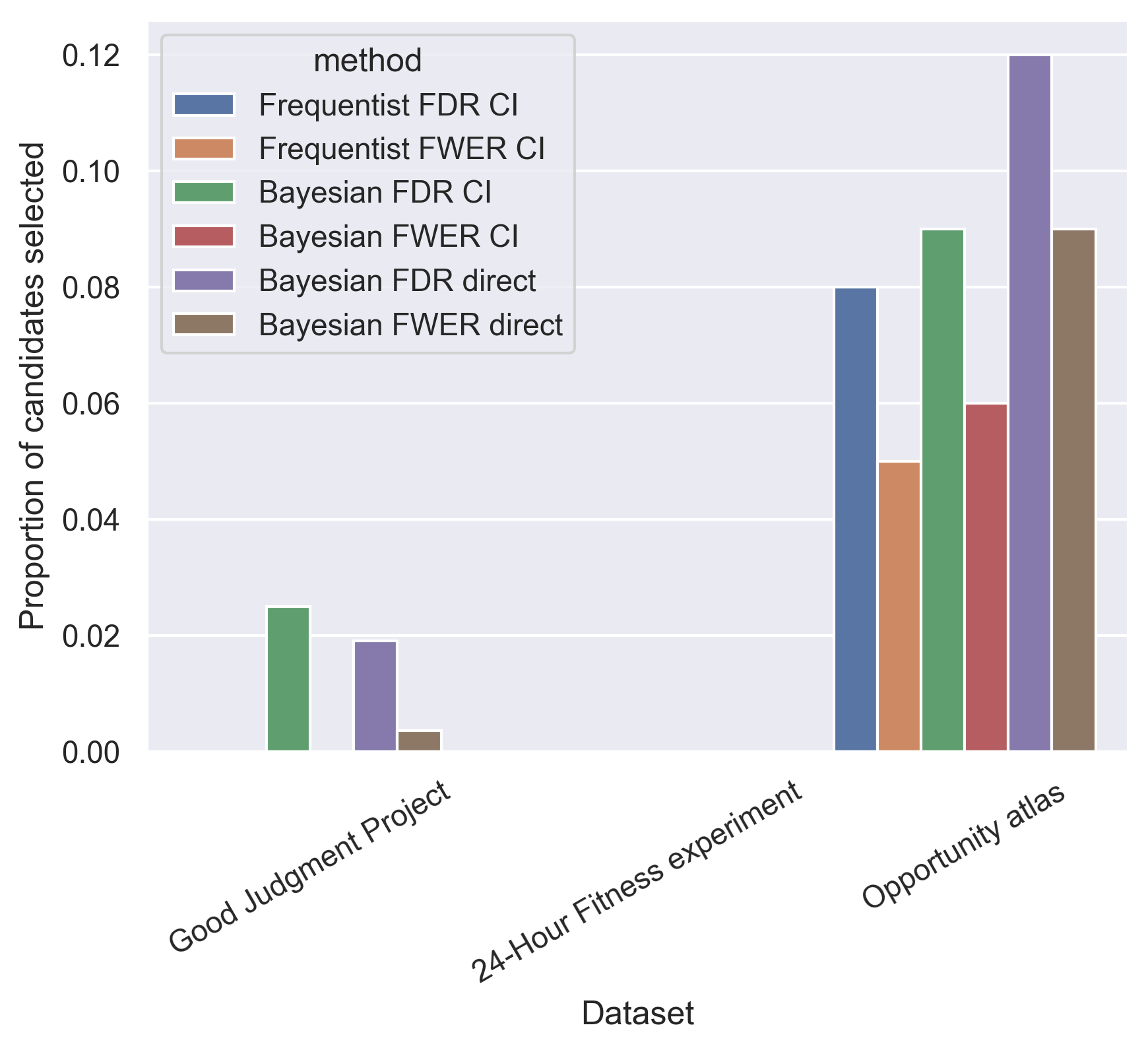}
    \caption{Proportion of candidates selected as being among the top 10\% with a 20\% error rate.}
    \label{fig:empirical_example_n_selected}
\end{figure}

For a more concrete illustration, Figure \ref{fig:empirical_example_n_selected} plots the results of these selection procedures for three datasets (the Good Judgment Project, 24-Hour Fitness, and Opportunity Altas datasets) when selecting the top 10\% of candidates with a 20\% error rate. For example, suppose we want to assemble a forecasting team in which 80\% of the team members are among the most accurate 10\% of forecasters. Using Bayesian selection and controlling the false discovery rate, we can select 2\% of forecasters from the Good Judgment Project as being among the most accurate 10\%. Alternatively, suppose we want to assemble a forecasting team which we are 80\% confident contains only forecasters among the most accurate 10\%. Using the Bayesian FWER direct method, we can select less than 1\% of forecasters. While it might seem like Bayesian procedures select very few forecasters, note that frequentist procedures select none.

Similarly, suppose we want to select the most effective behavioral interventions in the 24-Hour Fitness experiment for implementation or follow-up studies. Unfortunately, there is so much noise in the 24-Hour Fitness experiment that we cannot say which treatments are most effective when controlling the error rate at 20\%. All procedures select zero treatments as being among the most effective 10\%.

By contrast, there is much less noise in the Opportunity Atlas dataset. Using the Bayesian FDR direct method with a false discovery rate at 20\%, we select 12\% of commuting zones as being in the top 10\%. It may seem undesirable to say that 12\% of commuting zones are among the top 10\%. However, if the selected commuting zones are a superset of the true top 10\%, then the false discovery rate would be correct; $(0.12 - 0.10) / 0.12 = 0.17 < 0.2$. Alternatively, using the Bayesian FWER direct method with a family-wise error rate of 20\%, we select 9\% of commuting zones as being in the top 10\%. Note that this is a substantial improvement over the Bayesian SRCI method (which selects 6\%) and the frequentist SRCI method (which selects 5\%).

In sum, our selection procedures show similar results when applied directly to conventional estimates as they did in our simulations. Bayesian procedures select more candidates than their frequentist counterparts, with direct Bayesian selection performing exceptionally well.

\section{Conclusion} \label{sec:conclusion}

Decision-making often involves ranking and selection. Unfortunately, we rarely know candidates' true ranks and must perform ranking and selection based on noisy estimates. This paper developed new Bayesian algorithms for ranking and selection and showed that they outperform frequentist analogues using simulations and empirical illustrations. Specifically, Bayesian RCIs are shorter than frequentist RCIs and make only minimal coverage sacrifices, and Bayesian selection algorithms select more candidates than frequentist algorithms while controlling the error rates. We highlighted the practical applicability of our ranking and selection methods for field experiments, economic mobility analysis, and forecasting.

\clearpage

\singlespacing
\bibliographystyle{unsrtnat}
\bibliography{bibliography}

\end{document}